# On advanced fluid modelling of drift wave turbulence

J. Weiland[1], A. Zagorodny[2] and V. Zasenko[2]

[1] Chalmers University of Technology and EURATOM-VR Association
    Göteborg, Sweden.
[2] Bogoliubov Institute for Theoretical Physics, 03143 Kiev, Ukraine

## Abstract

The Dupree-Weinstock renormalization is used to prove that a reactive closure exists for drift wave turbulence in magnetized plasmas. The result is used to explain recent results in gyrokinetic simulations and is also related to the Mattor-Parker closure. The level of closure is found in terms of applied external sources.

A major complication in the description of drift-wave turbulence in magnetic confinement systems is that phase velocities of the turbulent perturbations are comparable to thermal velocities of the plasma[1]. Although this is true both parallel and perpendicularly to the magnetic field, the perpendicular (magnetic drift) resonance has turned out to play a particularly important role in particular in H-mode plasmas with flat density profiles[2]. Although the development of nonlinear gyrokinetics[3,4] has been rapid, it is still not feasible to use gyrokinetic codes as transport codes because of the computation time needed. Thus for predictive transport simulations we still need fluid models. In order to deal with the wave particle resonances, advanced fluid models[5-7] were developed in the end of the 1980's and 1990's. We will here define advanced fluid models as fluid models which give a rule for treating the wave particle resonances in such a way that the model can be used near these resonances also in collisionless plasmas. The first model designed to do this was the advanced reactive fluid model[5] and later the gyro fluid models in the US followed[6,7]. The improvement of the description of tokamak transport was dramatic[8]. The most significant feature was that a new regime, the 'flat density regime', which typically persists in the major part of L-mode plasmas and all the way out to the edge pedestal in H-mode plasmas, could now be described and the problem with the radial variation of drift wave transport coefficients was resolved[9].

Thus a general feature of all advanced fluid models is that the description of experimental results is much better than with previous standard fluid models. The main difference between the reactive model and gyro-fluid models is that gyro-fluid models add linear dissipative kinetic resonances thus getting a dissipative closure. Such models originally got too large transport as compared to gyro-kinetic models. The main reference is here from the Cyclone project[10]. In comparisons with gyro-kinetic models the question of which is the better closure obviously depends on to which extent nonlinearities in velocity space will modify or remove the dissipative linear kinetic resonances. Here, clearly, it is essential to keep velocity space nonlinearities[4]. These are often neglected in nonlinear Vlasov codes with the argument that they are small. When these are ignored the difference to quasilinear codes is usually rather small. A simple example of when linear resonances are completely removed due to flattening in velocity space is the bump in tail instability[11]. In this case the bump is removed completely due to simple quasilinear diffusion in velocity space. It seems unlikely that simple quasilinear diffusion could give complete flattening of the bulk distribution since the whole distribution can not be flattened and since a local flattening would give steeper gradients for neighboring velocities. It has recently been pointed out, however, that a complete flattening is not needed to remove dissipative resonances which are due only to moments higher than those included self-consistently[8]. This is because the full kinetic resonance involves all fluid resonances while in advanced fluid models the lowest order fluid resonances are kept unexpanded and treated self consistently. A significant effect, in kinetic theory, is, however played by the nonlinear frequency shift. It may shift the mode frequency between regions with negative and positive wave energy in such a way that dissipative wave particle resonances change sign. Thus the dissipative kinetic resonance can be averaged out in a way very similar to particle trapping without any change in the distribution function. This was actually the reason for the very strong difference between the linear and nonlinear closures in the Mattor-Parker work[12]. The Mattor-Parker system was later generalized to include effects of background turbulence through a diffusion term.

In the present paper we will generalize the Mattor-Parker work to include the kinetic nonlinearities and to show that wave-particle interaction can be seen as a collision between wave and particle where both change their velocity so that overall momentum is conserved. The change in wave phase velocity is due to a nonlinear frequency shift. We will also show that a renormalization leads to a situation where a reactive closure is valid, In such a state energy is clearly conserved.

In order to have conservation of energy, a kinetic code must include the nonlinear response to linear wave-particle resonant interaction. This has been generally done in Particle In Cell (PIC) codes[13-15] but not always in Vlasov codes[16,17].
Since Finite Larmor Radius (FLR) effects are not an essential part of the following discussion we will here start from the drift-kinetic equation.

$$\frac{\partial f}{\partial t} + (\hat{\mathbf{e}}_\| v_\| + \mathbf{v}_D + \mathbf{v}_E) \cdot \nabla f - \frac{q}{m} \nabla_\| f \frac{\partial f}{\partial v_\|} - \frac{q}{m} \nabla f \cdot \left( \frac{\mathbf{v}_{\nabla B}}{B} \frac{\partial f}{\partial \mathbf{m}} + \frac{\mathbf{v}_k}{v_\|} \frac{\partial f}{\partial v_\|} \right) = 0 \qquad (1)$$

**We here used standard notations.**
The term $\mathbf{v}_E \cdot \nabla f$ is the *ExB nonlinearity* which is of fluid type while the term $\nabla_\| f \frac{\partial f}{\partial v_\|}$

is the *parallel nonlinearity* and the last term, containing curvature and grad B parts, is the *magnetic drift nonlinearity*. Since the last two nonlinearities involve gradients in velocity space they will be called *kinetic nonlinearities*.
We can conveniently relate the nonlinearities through their bounce frequencies[18]:

$$w_B = \mathbf{k} \cdot \mathbf{v}_E; \quad w_\| = k_\| v_t \sqrt{ef/T_e}; \quad w_\perp = w_D \sqrt{ef/T_e} \qquad (2)$$

Since typically $k_\| v_t$ and $w_D$ are comparable, the kinetic nonlinearities will here be considered to be comparable.
Clearly the ratio of kinetic to ExB bounce frequencies can then be written:

$$\frac{w_\perp}{w_B} \approx \sqrt{\frac{w_D}{g}} \frac{1}{\sqrt{k_\perp R}} \approx \frac{1}{\sqrt{k_\perp R}}$$

which is of order $(?*)^{1/2}$, $?* = ?/a$. The smallness of the kinetic nonlinearities has often been taken as an argument for ignoring these in nonlinear kinetic codes. However, as we will see, they are only smaller than the ExB nonlinearity in the coherent case. Taking the parallel nonlinearity as a typical kinetic nonlinearity we can write the quasilinear diffusivity in velocity space as[11]:

$$D^v = p \sum \left| \frac{q f_k}{T} \right|^2 v_{th}^2 \frac{k_\|^2 v_{th}^2}{w} ; [m^2/s^3] \qquad (3a)$$

The energy diffusivity in real space can be estimated as[8]:

$$D = D_B \frac{\omega k_y c_s}{\omega^2 + \gamma^2} k_y r_s \left|\frac{q\phi}{T}\right|^2 \tag{3b}$$

The timescales of these processes then compare as:

$$\frac{df_0}{dt} = (D^v/v_{th}^2) f_0 \quad \text{and} \quad \frac{df_0}{dt} = (D/L^2) f_0$$

where the derivative in velocity space was replaced by $1/v_{th}$ and the derivative in real space by $1/L$ where L is a typical equilibrium inhomogeneity space scale. Then using the diffusivities in velocity and real space according to (3), using $k_\| v_{th} \approx \omega$ we get the ratio:

$$\frac{D^v}{v_{th}^2} / \frac{D}{L^2} = \rho \frac{k_y^2 D_B}{k_y c_s k_y r_s} = \rho \propto 1 \tag{4}$$

Where we estimated the frequency by the drift frequency $\quad \omega = \dfrac{k_y r_s c_s}{L} = D_B k_y / L$

using the same equilibrium scale length as before. Thus *the time scales of diffusion in real and velocity spaces are comparable*! This means that the kinetic nonlinearities have to be included in kinetic codes which operate on the time scale of confinement in real space. This is, on the other hand, necessary for a code that aims at describing transport in real space. We can also draw another interesting conclusion from this. In the stationary, phase mixed, turbulent state, the nonlinearities only give transport. Thus, since transport in real and velocity space is comparable, all the nonlinearities are comparable in the final phase mixed state!

So far we have only considered the quasilinear velocity space diffusion. Although this may not be sufficient for creating a reactive closure in the bulk of the velocity distribution, it represents a main mechanism which will be an important part of the full dynamics and it gives the correct timescale for turbulent wave-particle interactions. In general we will, however, have to go beyond the quasilinear description. We will then have to include nonlinear frequency shifts into our description. In order to see how nonlinear frequency shifts will influence the dynamics we will first go to the comparatively simple coherent (strongly nonlinear) limit. In this limit Mattor and Parker[12] derived a nonlinear three-wave system for slab ITG modes corresponding to the generation of zonal flows by self-interaction. In this system a nonlinear closure was introduced for the heat flow so that the the third moment was described kinetically. This lead to a plasma dispersion function including a nonlinear frequency shift. This nonlinear closure turned out to be vastly superior to the Hammett-Perkins[6] closure in comparison with a drift kinetic model. A simple example on the potential importance of nonlinear frequency shifts is the growthrate of the universal drift wave instability. It can be written:

$$g = \left(\frac{\pi}{2}\right)^{1/2} \omega_{*e} \frac{\omega - \omega_{*e}}{k_\parallel v_{te}} e^{-\omega^2/(k_\parallel v_{te})} \quad (5)$$

We here have a kinetic instability (inverse electron Landaudamping) when $\omega > \omega_{*e}$ although the velocity distribution is Maxwellian. This is because of the space inhomogeneity which makes the wave energy negative. A nonlinear frequency shift into the region with positive energy reverses the sign of the growth-rate. This happens in the Mattor Parker system for self-interaction of a slab ITG mode generating a zonal flow.
This system was later generalized to a closure two orders higher (fifth moment kinetic) with inclusion of diffusion due to background turbulence[19]. The zonal flow mode had zero poloidal modenumber and a second harmonic radial modenumber. The resulting time evolution is shown in Fig 1.

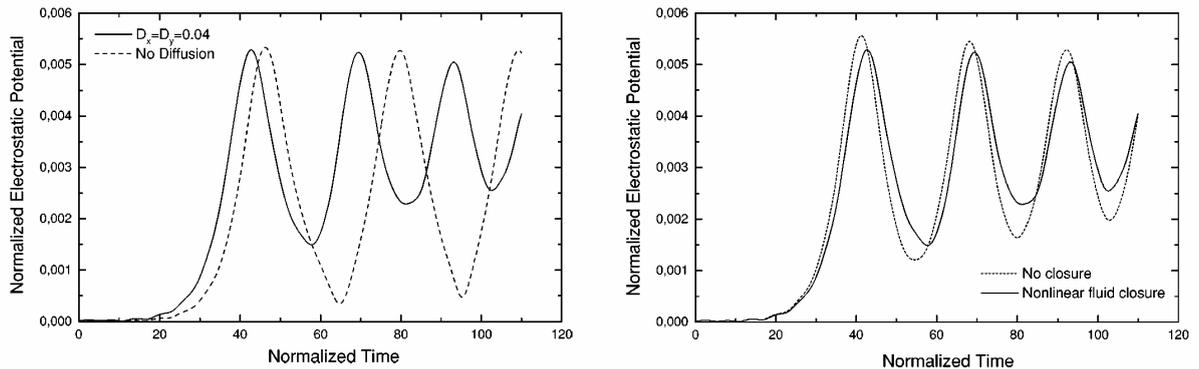

a) Time evolution of norm. density with and without diffusion    b) The same evolution as in a) with and without closure

Fig 1. Time evolution of normalized density in a system of zonal flow generation by self-interaction of a wave with nonlinear fluid closure. . Here $k_x = k_y = 0.3$, $\eta_i = 3$   This figure has been taken from: I. Holod, J. Weiland and A. Zagorodny, Phys. Plasmas **9**, 1217 (2002). With the permission of the American Institute of Physics.

We here note how well the reactive closure (b without closure) is doing. The diffusion causes relaxation towards an asymptotic equilibrium. This process would speed up and the equilibrium would be lower with absorbing boundaries in k-space.
The diffusion here corresponds to the effect of background turbulence. This generalizes the present system to a partially coherent system and the asymptotic equilibrium state will correspond to a weakly turbulent state. In the present system the parallel nonlinearity was not included. According to our previous discussion this nonlinearity will become important after sufficient phase mixing of the ExB nonlinearity.

We now return to the kinetic description of a phase mixed turbulent state. We have here, so far, only discussed the quasilinear limit which just includes diffusion due to independent (no mode coupling) linear waves. As noted above such a simple system may not be able to give a nonlinear state where a reactive closure is valid. We will now apply the Dupree-Weinstock renormalization[20,21]. It means including the phase coherent nonlinear couplings on the drift waves which then become coupled.

$$i(\mathbf{k} \cdot \mathbf{v} - w_k)f_k + \frac{q}{m}\mathbf{E} \cdot \frac{\partial f_0}{\partial \mathbf{v}} = -\frac{q}{m}\sum \mathbf{E}_{k-k'} \cdot \frac{\partial f_{k'}}{\partial \mathbf{v}} e^{i\Delta w_{k,k'}t} \tag{6}$$

By solving the corresponding equation for $f_{k'}$ and substituting this into (6) we get cubic nonlinear terms. The only phase coherent terms are here those that have the form of an intensity multiplying $f_{k'}$. In a phase-mixed turbulent state, the remaining couplings will phase mix to zero and only the phase coherent terms will remain. If we substitute also the linear acceleration term into the convolution, replace one E by the zero frequency nonlinearly driven (ponderomotive) E and the other by f using the linear relation we get a resonant term. The velocity derivative here gives a velocity and we have obtained the friction term in a Fokker-Planck equation. A more general, systematic derivation is given in Ref 22. We write the Fokker Planck equation in real space as:

$$(\frac{\partial}{\partial t} + v\frac{\partial}{\partial x})f(x,v,t) = \frac{\partial}{\partial v}\left[\boldsymbol{b}v + D^v\frac{\partial}{\partial v}\right]f(x,v,t) \tag{7}$$

It has the exact analytical solution[24]: 
$$f(x,v,t) = \frac{e^{bt}}{2p\Delta^{1/2}}e^{\left[-\frac{1}{2\Delta}(a\boldsymbol{dr}^2 + 2h\boldsymbol{dr}\boldsymbol{dP} + b\boldsymbol{dP}^2)\right]} \tag{8}$$

Where:

$$a = \frac{2}{b^2}D^v t \qquad \boldsymbol{dr} = \mathbf{v}e^{bt} - \mathbf{v}' \qquad b = \frac{1}{b}D^v(e^{2bt}-1) \qquad h = -\frac{2}{b^2}D^v(e^{bt}-1)$$

$$\boldsymbol{dP} = x + \Delta v/\boldsymbol{b} \qquad \Delta = ab - h^2 \qquad \boldsymbol{t} = t - t'$$

Here t is the present time and t' is a given time where the state is known. Index ' refers to the time t' for all variables

By using (8) as a weight function for ensemble averages we obtain the mean square velocity deviation ($\boldsymbol{dv} = \mathbf{v} - \mathbf{v}'$):

$$\langle dv^2 \rangle = \frac{D^v}{b}(1 - e^{-2bt}) + v'^2(1 - e^{-bt})^2 \tag{9a}$$

With the asymptotic limit: 
$$\langle dv^2 \rangle = \frac{D^v}{b} + v'^2 \leftrightarrow (\boldsymbol{t} \to \infty) \tag{9b}$$

This is a very important property. It means that asymptotically, after a time $t \approx 1/(2b)$ the mean square velocity deviation will be constant. Thus after this time there will be *no more energy transfer between waves and particles on the average*. Thus dissipative kinetic resonances are averaged out and *a reactive fluid closure must be possible*!

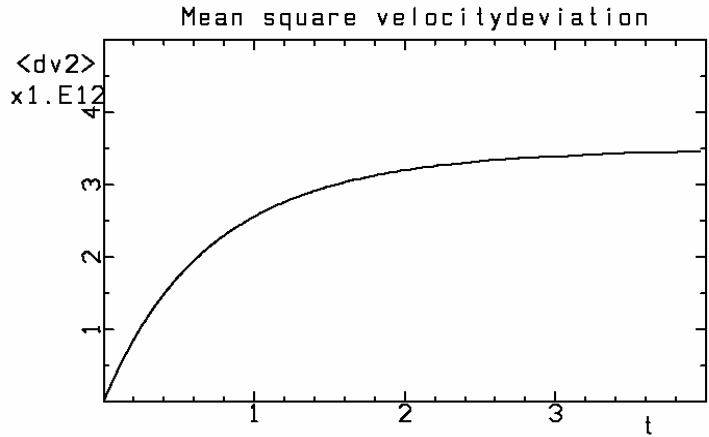

Fig2. Mean square velocitydeviation, <dv2>, as a solution of (9a) with ß =1, $D^v = 2.5 \, 10^{12}$ and v'=$10^6$.

We have here added a strongly nonlinear feature corresponding to a finite trapping time. Thus we are, in fact, considering effects of trapping! It is, however, very important to realize that we picked out phase coherent terms in a situation with as random phases as possible. Thus the strongly nonlinear effects we included here can never be avoided. As it turns out, the longest possible timescale for the turbulent friction to act on is the confinement timescale. These effects are clearly secular (because the velocity space nonlinearities are all the time directed so as to take particles out of resonance) so the mean square velocity deviation is bound to saturate at the latest on the confinement timescale. The qualitative result in Fig2, obtained from the analytical solution (9) has also been recovered in several direct particle simulations, including all nonlinear terms, as shown in Ref's 22, 23 and 25. In more coherent situations the saturation will happen earlier. Such a case was found[25] in a simulation of the beam-plasma instability in a situation with Kubo number (ratio of trapping and correlation times) close to 1. There a saturation time consistent with trapping between wave packets was found. Thus the validity of the analytical solution (9) has been tested by computer simulations in various regimes. In order to get higher order phase coherent contributions we will have to go to fifth order nonlinearities. Since the small parameter in this expansion is of order $10^{-2}$, such terms will be ignorable.

Since we, in the renormalized system, have included trapping on a long time scale, we might think that the kinetic resonance has been averaged out completely. However, the kinetic resonance is the sum of all fluid resonances and fluid resonances associated with fluid moments that have external sources can still be maintained. The fact that we now have a reactive system actually only means that we now have a finite number of fluid resonances. Since the force towards thermodynamic equilibrium tends to make all fluid moments without sources homogeneous at the value at the outer boundary, only moments with sources will remain. We furthermore know empirically that we need *external* sources for both density and temperature although density gradients may drive turbulent temperature pinches and vice versa. Thus turbulent pinches need external sources to feed them. Collisions can transfer an external current source to heat but this is a completely random process so collisions can not give a source for higher moments which require correlated sources (compare the situation for the irreducible fourth moment in Ref 26). Furthermore we realize that adding collisions to the Fokker-Planck equation would only speed up the approach to the asymptotic state. Clearly we need to keep self-consistently all moments with external sources in the experiment and moments that can be expressed directly in these like the diamagnetic heat flow. We also conclude that a nonlinear kinetic code will not have reached steady state until the mean squared velocity deviation is constant in time, that is when it has reached stationary state in velocity space.

We have here shown that a turbulent plasma will always approach a state where a reactive fluid closure is valid after a sufficiently long time. This timescale can not be longer than of the order of the confinement time in real space. On a short timescale after the initial saturation of the most unstable modes, the ExB nonlinearity dominates. It tends to average out kinetic resonances by nonlinear frequency shifts, changing the sign of the wave energy. On a longer timescale the kinetic nonlinearities will lead to a state where there is no more energy exchange between waves and particles on the average. This is an exact analytical result supported also by direct particle simulations. We point out here that Ref 23 actually contains the full system, i.e. derives and simulates the sationary asymptotic state for the mean square velocity deviation with the full resonance, including also the magnetic drift. In this state energy will be conserved for the turbulence and the resonant particles separately (no exchange) and as is well known energy conservation requires the inclusion of kinetic nonlinearities in kinetic codes. As has been seen both in GTC simulations[14,15] including the parallel nonlinearity and in analytic calculations[27], zonal flows are significantly stronger and transport is reduced in situations where dissipative kinetic resonances are reduced. Thus an effect of the kinetic nonlinearities is definitely expected at the latest at times comparable to the confinement time. Actually, the time needed for the parallel nonlinearity to become important can be shorter due to efficient phase mixing of the ExB nonlinearity.